\documentstyle[epsbox,epsf]{llncs}

\begin{document}
%\draft
\title{A Simple Perceptron that \\ Learns  
Non-Monotonic Rules\footnote[4]{To appear in Hong Kong 
International Workshop on 
Theoretical Aspects of Neural Computation (TANC-97): 
A Multidisciplinary Perspective}}
\author{Jun-ichi INOUE\inst{1}\inst{3}\footnote[7]{E-mail address: 
jinoue@stat.phys.titech.ac.jp or jinoue@zoo.riken.go.jp}, Hidetoshi NISHIMORI\inst{1} 
and Yoshiyuki KABASHIMA\inst{2}}
\institute{Department of Physics,  
Tokyo Institute of Technology, \\
Oh-okayama, Meguro-ku, 
Tokyo 152, Japan 
\and
Department of 
Computational Intelligence and 
Systems Science, \\
Interdisciplinary 
Graduate School of Science 
and Engineering, \\ 
Tokyo Institute 
of Technology, 
Yokohama 226, Japan
\and 
Laboratory  for  Information Representation, 
RIKEN, 
Hirosawa 2-1, Wako-shi, Saitama 351-01, Japan
}
\maketitle
%%%%%%%%%%%%%%%%%%%%%%%%%%%%%%%%%%%%%%%%%%%%%%%%%%%%%%%%%%%%
%%                                                        %%
%%                    Abstruct                            %%
%%                                                        %%
%%%%%%%%%%%%%%%%%%%%%%%%%%%%%%%%%%%%%%%%%%%%%%%%%%%%%%%%%%%%
%%
%%
\begin{abstract}
We investigate the generalization ability 
of a simple perceptron trained 
in the off-line and on-line supervised modes. 
Examples are extracted from the teacher 
who is a non-monotonic perceptron. 
For this system, difficulties of training  
can be controlled continuously 
by changing a parameter of the teacher.  
We train the student by several learning 
strategies in order to obtain the theoretical 
lower bounds of generalization errors under various conditions. 
Asymptotic behavior of the learning curve has been derived, 
which enables us to 
determine the most suitable 
learning algorithm 
for a given value of 
the parameter controlling difficulties 
of training. 
\end{abstract}
%%
%%
%%
%%%%%%%%%%%%%%%%%%%%%%%%%%%%%%%%%%%%%%%%%%%%%%%%%%%%%%%%%%
%%%%%%%%%%%%%%%%%%%%%%%%%%%%%%%%%%%%%%%%%%%%%%%%%%%%%%%%%%
\section{Introduction}
%%%%%%%%%%%%%%%%%%%%%%%%%%%%%%%%%%%%%%%%%%%%%%%%%%%%%%%%%%
%%%%%%%%%%%%%%%%%%%%%%%%%%%%%%%%%%%%%%%%%%%%%%%%%%%%%%%%%%
%%
%%
%%
Learning from examples has been one of the most 
attractive problems for computational 
neuroscientists 
\cite{Hertz,Watkin93,Opper95,Amari,Amari93,Amari94,Boes93,Kaba94,Bouten,Broeck}. 
For a given system, 
superiority of the learning 
strategy should be measured by the 
generalization error, namely the probability 
of disagreement between the teacher and student outputs 
for a new example after the student has been trained. 
Much efforts have been 
invested into 
investigations in the case of 
learnable rules, 
and it is desirable to 
construct suitable 
learning strategies and minimize 
the residual generalization error 
even if it is impossible for the student 
to reproduce the teacher  
input-output relations perfectly. 
In the present contribution we 
investigate the generalization error for such an 
unlearnable case \cite{Watkin92,Kaba2,Kim,Saad,Kaba,Inoue1,Inoue2}. 

In our model system, the student is a 
simple perceptron whose output is given as 
$S(u)={\rm sign}(u)$ with 
$u{\equiv}\sqrt{N}({\bf J}{\cdot}{\bf x})/|{\bf J}|$, 
where ${\bf J}$ is the synaptic weight vector and 
${\bf x}$ is 
a random input vector which is extracted from 
the $N$-dimensional sphere $|{\bf x}|^{2}=1$.
The teacher is a non-monotonic (or reversed-wedge type) 
perceptron whose output is 
represented as $T_{a}(v)={\rm sign}[v(a-v)(a+v)]$ with 
$v{\equiv}\sqrt{N}({\bf J}^{0}{\cdot}{\bf x})$. 
The weight vector of the teacher has been written as 
${\bf J}^{0}$.
If $a=0$ or $a=\infty$, the student 
can learn the teacher rule perfectly, 
the learnable case. 

If the width $a$ of the reversed wedge 
is finite, the student 
can not reproduce 
the teacher input-output relations 
perfectly and the generalization error 
remains non-vanishing even after 
infinite number of examples have been presented. 
For this system, when the overlap between 
the teacher and student 
is written as $R\,{\equiv}\,
({\bf J}{\cdot}{\bf J}^{0})/|{\bf J}||{\bf J}^{0}|$, 
the generalization error ${\epsilon}_{g}$ 
is 
\begin{eqnarray}
{\epsilon}_{g}\,{\equiv}\,{\ll}
{\Theta}(-T_{a}(v)S(u)){\gg} \hspace{2.1in}\nonumber \\
\mbox{}=2\int_{a}^{\infty}Dv\,H
\left(
-\frac{Rv}{\sqrt{1-R^{2}}}
\right)
+2\int_{0}^{a}Dv\, H
\left(
\frac{Rv}{\sqrt{1-R^{2}}}
\right) \nonumber \\ 
\mbox{}{\equiv}\,E(R ),  \hspace{2.95in} 
\end{eqnarray}
where $H(x)=\int_{x}^{\infty}Dt$ 
with $Dt\,{\equiv}\,
{\exp}(-t^{2}/2)/\sqrt{2\pi}$ 
and ${\ll}{\cdots}{\gg}$ stands for the averaging over the 
connected Gaussian distribution: 
\begin{equation}
P_{R}(u,v)=\frac{1}{2\pi\sqrt{1-R^{2}}}{\exp}\left[
-\frac{(u^{2}+v^{2}-2Ruv)}{\sqrt{2(1-R^{2})}}
\right].
\end{equation}
It is important that 
this expression 
is independent of 
specific learning algorithms. 
In Fig. 1 we plot  $E(R)$ for several values of 
$a$. 
\begin{figure}
\begin{center}
\psbox[width=8cm]{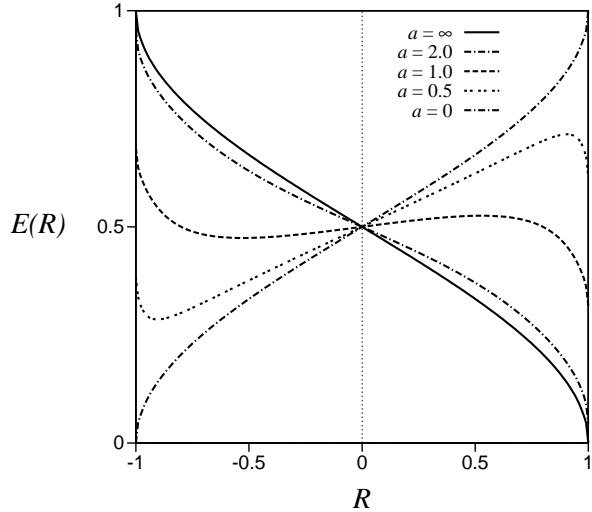}
\end{center}
\caption{
Generalization error as a function of $R$ for 
$a=\infty$, $2$, $1$, $0.5$ and $a=0$.
}
\end{figure}

Minimization of $E(R)$ with respect to 
$R$ gives the theoretical lower bound of 
the generalization error. 
In Fig. 2 we show the theoretical 
lower bound corresponding to the minimum 
value of $E(R)$ in Fig. 1 and in 
Fig. 3  we plot  
the corresponding optimal overlap $R_{\rm opt}$ which gives 
the bound. 
\begin{figure}
\begin{center}
\psbox[width=8cm]{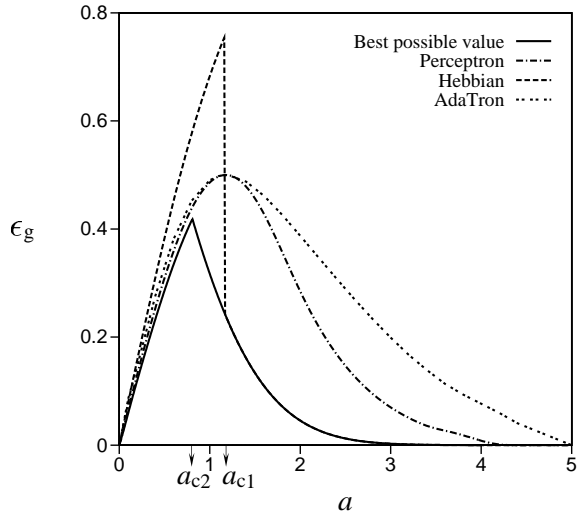}
\end{center}
\caption{The best possible value (theoretical lower bound) of the 
generalization error, 
the residual generalization errors of 
conventional Hebbian, perceptron  
and AdaTron learning algorithms are plotted as functions of $a$. 
}
\end{figure}
\begin{figure}
\begin{center}
\psbox[width=8cm]{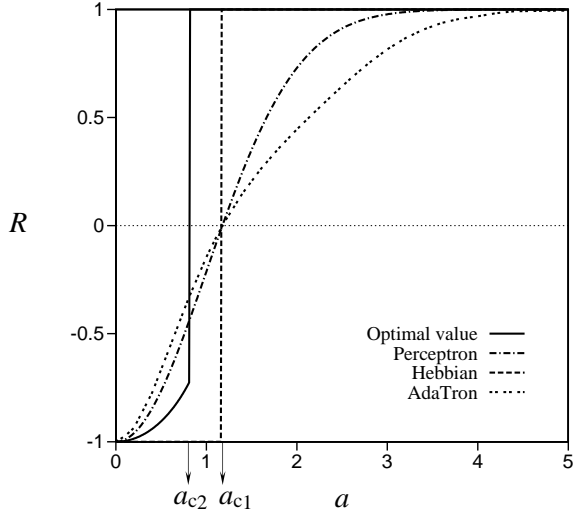}
\end{center}
\caption{
The optimal overlap $R$ which gives the best possible 
value and overlaps which give the residual errors in Fig. 2  
for Hebbian, perceptron and AdaTron learning algorithms.
}
\end{figure}

From Fig. 3 we see that 
one should train the student 
so that $R$ becomes $1$ for $a>a_{c2}=0.80$.
For $a<a_{c2}=0.80$,  the optimal $R$ is not $1$ 
but $R_{*}=-\sqrt{(2{\log}2-a^{2})/2{\log}2}$. 
This system 
shows the first order phase transition at $a=a_{c2}$ 
and the optimal overlap changes from $1$ to $R_{*}$ 
discontinuously. 

In the following sections, 
we investigate 
various learning 
strategies to clarify 
the asymptotic behavior of learning curves.
%%
%%
%%
%%%%%%%%%%%%%%%%%%%%%%%%%%%%%%%%%%%%%%%%%%%%%%%%%%%%%%%%%%%%%%%%%
%%%%%%%%%%%%%%%%%%%%%%%%%%%%%%%%%%%%%%%%%%%%%%%%%%%%%%%%%%%%%%%%%
\section{Off-line learning}
%%%%%%%%%%%%%%%%%%%%%%%%%%%%%%%%%%%%%%%%%%%%%%%%%%%%%%%%%%%%%%%%%
%%%%%%%%%%%%%%%%%%%%%%%%%%%%%%%%%%%%%%%%%%%%%%%%%%%%%%%%%%%%%%%%%
%%%%%%%%%%%%%%%%%%%%%%%%%%%%%%%%%%%%%%%%%%%%%%%%%%%%%%%%%%%%
%%            Section of off-line learning                %%
%%%%%%%%%%%%%%%%%%%%%%%%%%%%%%%%%%%%%%%%%%%%%%%%%%%%%%%%%%%%
%%
%%
We first investigate the 
generalization ability of the student in 
off-line (or batch) mode 
following the minimum error algorithm. 
The minimum error algorithm 
is a natural learning strategy to minimize 
the total error for $P$ sets of 
examples $\{{\bf \xi}^{P}\}$
\begin{equation}
E({\bf J}|\{{\bf \xi}^{P}\})=\sum_{\mu=1}^{P}{\Theta}
(-T_{a}^{\mu}{\cdot}u^{\mu})
\label{terror}
\end{equation}
where we set $u^{\mu}\,{\equiv}\,({\bf J}{\cdot}{\bf x}^{\mu})/\sqrt{N}$. 
From the energy defined by Eq. (\ref{terror}), the 
partition function  with the inverse temperature $\beta$ is given by 
\begin{eqnarray}
Z(\beta)=\int{d}{\bf J}\,{\delta}
(|{\bf J}|^{2}-N)
{\exp}\left(-{\beta}E({\bf J}|\{{\xi}^{P}\})\right) \nonumber \\
\mbox{}=\int{d}{\bf J}\,{\delta}
(|{\bf J}|^{2}-N)\displaystyle{\prod_{\mu=1}^P}\left[
{\rm e}^{-\beta}+(1-{\rm e}^{-\beta})
{\Theta}(-T_{a}^{\mu}{\cdot}u^{\mu})
\right]
\label{fpart}
\end{eqnarray}
There exists weight vectors 
that  reproduce 
input-output relations completely if ${\alpha}=P/N$ 
is smaller than a  critical capacity ${\alpha}_{c}$. 
Therefore, we can calculate the learning 
curve (LC) below ${\alpha}_{c}$ by evaluating 
the logarithm of the Gardner-Derrida volume 
$V_{\rm GD}=Z(\infty)$ as 
\begin{equation}
\frac{{\log}V_{\rm GD}}{N}=\frac{{\ll}{\log}Z(\infty){\gg}_{\{{\xi}^{P}\}}}
{N}=\frac{1}{N}\displaystyle{\lim_{n{\rightarrow}0}}
\frac{{\ll}Z^{n}(\infty){\gg}_{\{{\xi}^{P}\}}-1}
{n}.
\label{repli1}
\end{equation}
On the other hand, at ${\alpha}={\alpha}_{c}$, 
$V_{\rm GD}$ shrinks to zero and for 
${\alpha}>{\alpha}_{c}$, 
we can not find the solution in the weight space. 
Then, we treat the next free energy 
\begin{equation}
-f=\displaystyle{\lim_{{\beta}{\rightarrow}\infty}}
\frac{{\ll}{\log}Z(\beta){\gg}_{\{{\xi}^{P}\}}}
{N{\beta}}=
\displaystyle{\lim_{\beta{\rightarrow}\infty}}\,
\displaystyle{\lim_{n{\rightarrow}0}}
\frac{{\ll}Z^{n}(\beta){\gg}_{\{{\xi}^{P}\}}-1}{N{\beta}n}
\label{repli2}
\end{equation}
to find the solution weight  
${\bf J}$ which gives a minimum error for ${\alpha}>{\alpha}_{c}$. 
Introducing the order parameters 
$R_{\alpha}=({\bf J}^{0}{\cdot}{\bf J}_{\alpha})/N$ and 
$q_{\alpha\beta}=({\bf J}_{\alpha}{\cdot}{\bf J}_{\beta})/N$ and 
using the replica symmetric approximation 
$R_{\alpha}=R$ and $q_{\alpha\beta}=q$, 
Eq. (\ref{repli1}) is evaluated as 
\begin{equation}
{\rm ext}_{\{R,q\}}\left\{
2{\alpha}\int{Dt}\,{\Omega}(R/\sqrt{q}:t)\,
{\log}\,{\Xi}(q:t)+\frac{1}{2}{\log}(1-q)
+\frac{q-R^{2}}{2(1-q)}
\right\}
\label{ext1}
\end{equation}
with 
\begin{eqnarray}
{\Omega}(R:t)=\int{Dz}\,{\Big [}
{\Theta}(-z\sqrt{1-R^{2}}-Rt-a)
+{\Theta}(z\sqrt{1-R^{2}}+Rt) \nonumber \\
\mbox{}-{\Theta}(z\sqrt{1-R^{2}}+Rt-a)
{\Big ]},
\label{Omega}
\end{eqnarray}
\begin{equation}
{\Xi}(q:t)=\int{Dz}\,{\Theta}(z\sqrt{1-R^{2}}+t\sqrt{q}). 
\label{xi}
\end{equation}
And Eq. (\ref{repli2}) is evaluated as 
\begin{eqnarray}
{\rm ext}_{\{R,x\}}
{\bigg \{}
-2{\alpha}\left[
\int_{-\infty}^{0}{Dt}\,{\Omega}(R:t)
\left\{
{\Theta}(-t-\sqrt{2x})+\frac{t^{2}}{2x}
{\Theta}(t+\sqrt{2x})
\right\}
\right] \nonumber \\
\mbox{}+\frac{1-R^{2}}{2x}
{\bigg \}}
\label{ext2}
\end{eqnarray}
where we have set $x={\beta}(1-q)$ to find a non-trivial solution in 
the limit of ${\beta}{\rightarrow}\infty$ and $q{\rightarrow}1$. 
By solving the saddle point equation from Eqs. (\ref{ext1}) and 
(\ref{ext2}), we found that the LC is 
classified into the following five types depending on 
the parameter $a$. 
\begin{itemize}
\item 
$a=0,\infty$ (learnable case)\\ 
The solutions of  the saddle point equation are thermodynamically 
stable and the LC behaves asymptotically as \cite{Gyoe,Opper91} 
\begin{equation}
{\epsilon}_{g}\,{\sim}\,0.624\,{\alpha}^{-1}. 
\label{learnable}
\end{equation}
\item 
$a>a_{c0}\,{\sim}\,1.53$\\ 
The order parameter 
$R$ monotonically increases to $1$ as ${\alpha}
{\rightarrow}\infty$. 
The LC behaves asymptotically as 
\begin{equation}
{\epsilon}_{g}-{\epsilon}_{min}\,{\sim}\,{\alpha}^{-1}. 
\label{lc2}
\end{equation}
\item 
$a_{c0}>a>a_{c1}$\\ 
A first order phase transition 
from the poor generalization phase 
to the good generalization phase is 
observed at ${\alpha}\,{\sim}\,{\cal O}(1)$ in this 
parameter region (see Fig. 4). 
In the limit ${\alpha}{\rightarrow}\infty$, $R$ 
approaches to $1$ which achieves 
the global minimum of the 
generalization error in this 
parameter region and the asymptotic LC 
is identical to Eq. (\ref{lc2}). 
\item 
$a_{c1}>a>a_{c2}$\\  
The first order phase transition is observed 
similarly to the previous parameter region of $a$ (see Fig. 5).
However, the spinodal point ${\alpha}_{\rm sp}$ becomes 
infinity. 
The asymptotic form of the LC for 
this parameter region of $a$ 
is the same as Eq. (\ref{lc2}). 
\item 
$a_{c2}>a>0$\\ 
In this parameter region $E(R)$ is 
minimized not at $R=1$ but at $R=R_{*}$. 
Therefore, the solution $(R,x)=(R_{*},0)$ is the global 
minimum of the free energy for all values of $\alpha$ 
and there is no phase transition. 
The LC decays to its minimum as 
\begin{equation}
{\epsilon}_{g}-{\epsilon}_{min}\,{\sim}\,{\alpha}^{-2/3}.
\label{lc3}
\end{equation}
\end{itemize}
This result implies that the 
non-monotonic teacher with 
small $a$ is more difficult for a simple 
perceptron to learn than that 
with large $a$ \cite{Kaba}.
We conclude that minimum error algorithm can lead to the 
best possible value of the generalization error (see Fig. 2) for all 
values of $a$. 
Watkin and Rau \cite{Watkin92} also investigated the 
LC for the same system as ours, however, they 
investigated only ${\cal O}(1)$ range of ${\alpha}$. 
In this section, we investigated the LC 
for all ranges of ${\alpha}$.
\begin{figure}
\begin{center}
\psbox[width=8cm]{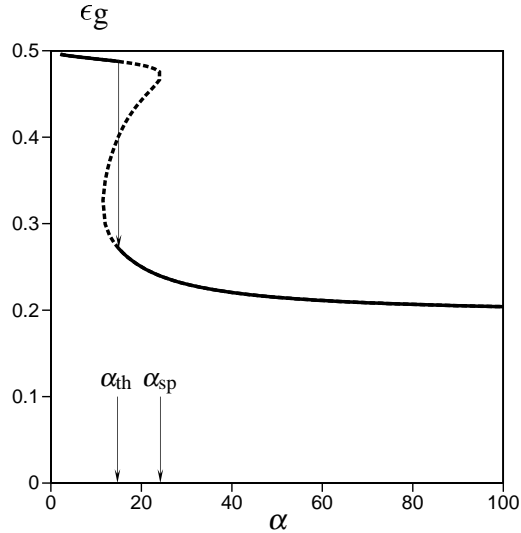}
\end{center}
\caption{
The learning curve  for the case of $a=1.3$. A first order phase transition 
appears at ${\alpha}_{\rm th}\,{\simeq}\,14.7$. The spinodal point is at 
${\alpha}_{\rm sp}\,{\simeq}\,24.2$.
}
\end{figure}
\begin{figure}
\begin{center}
\psbox[width=8cm]{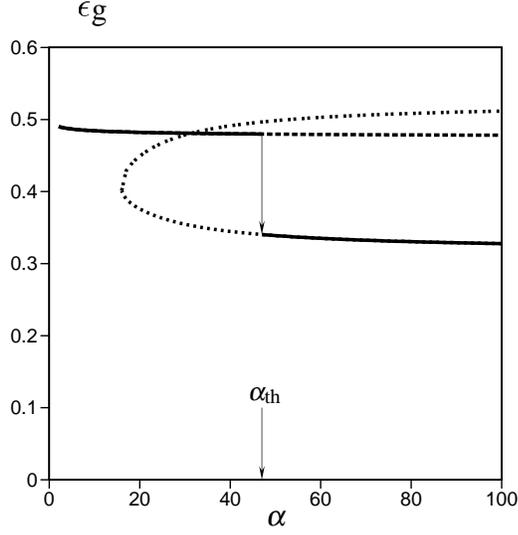}
\end{center}
\caption{
The learning curve  for the case of $a=1.0$. A first order phase transition 
appears at ${\alpha}_{\rm th}\,{\simeq}\,47$ and 
${\epsilon}_{g}$ changes discontinuously from the branch 1 to the branch 2. 
The spinodal point ${\alpha}_{\rm sp}$ has gone to infinity.
}
\end{figure} 
%%%%%%%%%%%%%%%%%%%%%%%%%%%%%%%%%%%%%%%%%%%%%%%%%%%%%%%%%%%%%%%%%
%%%%%%%%%%%%%%%%%%%%%%%%%%%%%%%%%%%%%%%%%%%%%%%%%%%%%%%%%%%%%%%%%
\section{On-line learning dynamics}
%%%%%%%%%%%%%%%%%%%%%%%%%%%%%%%%%%%%%%%%%%%%%%%%%%%%%%%%%%%%%%%%%
%%%%%%%%%%%%%%%%%%%%%%%%%%%%%%%%%%%%%%%%%%%%%%%%%%%%%%%%%%%%%%%%%
%%
%%%%%%%%%%%%%%%%%%%%%%%%%%%%%%%%%%%%%%%%%%%%%%%%%%%%%%%%%%%%%%%%%
\subsection{Conventional on-line learning algorithms}
%%%%%%%%%%%%%%%%%%%%%%%%%%%%%%%%%%%%%%%%%%%%%%%%%%%%%%%%%%%%%%%%%
%%
%%
The on-line learning dynamics 
we investigate in this work is 
generally written as follows. 
\begin{equation}
{\bf J}^{m+1}={\bf J}^{m}+gf(T_{a}(v),u)\,{\bf x}, 
\label{dynamic}
\end{equation}
where $m$ is the number of the presented patterns and 
$g$ is the learning rate.
In the limit of large $N$, the 
recursion 
relation Eq. (\ref{dynamic}) of the $N$-dimensional 
vector ${\bf J}^{m}$ is reduced to 
a set of differential equations for 
$R$ and $l=|{\bf J}|/\sqrt{N}$:  
\begin{equation}
\frac{dl}{d\alpha}=
\frac{1}{2l}{\ll}
g^{2}f^{2}(T_{a}(v),u)+2gf(T_{a}(v),u)ul
{\gg}
\end{equation}
\begin{equation}
\frac{dR}{d\alpha}=
\frac{1}{l^{2}}
{\ll}
-\frac{R}{2}g^{2}f^{2}(T_{a}(v),u)-
(Ru-v)gf(T_{a}(v),u)l
{\gg} 
\end{equation}
where 
${\alpha}$ is the number of presented patterns  per 
system size $m/N$.
In the present subsection we set $g=1$.
We now restrict ourselves to the following 
well-known algorithms:
\begin{itemize}
\item{Perceptron learning : 
$f=-S(u){\Theta}(-T_{a}(v)S(u))$}
\item{Hebbian learning : $f=-T_{a}(v)$}
\item{AdaTron learning : $f=-u\,{\Theta}\,(-T_{a}(v)S(u))$}.
\end{itemize} 
For the above three learning 
strategies, asymptotic forms of the 
generalization error for the learnable case are given as 
\cite{Watkin93,Opper95}: 
\begin{itemize} 
\item{Perceptron learning : ${\epsilon}_{g}\,{\sim}\,{\alpha}^{-1/3}$} 
\item{Hebbian learning : ${\epsilon}_{g}\,{\sim}\,{\alpha}^{-1/2}$} 
\item{AdaTron learning : ${\epsilon}_{g}\,{\sim}\,{\alpha}^{-1}$}.
\end{itemize}
\begin{figure}
\begin{center}
\psbox[width=8cm]{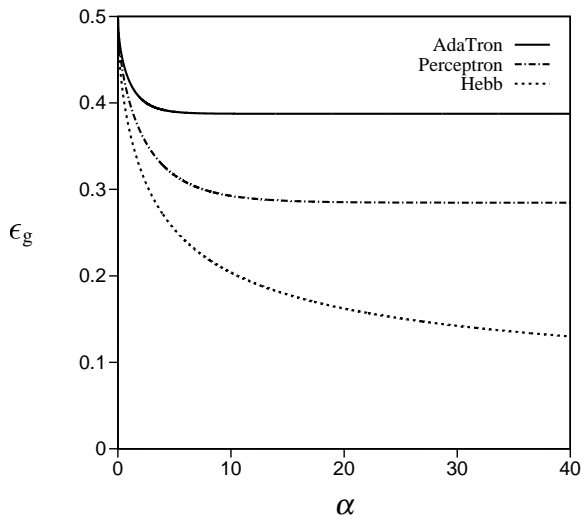}
\end{center}
\caption{
Generalization errors of the AdaTron, perceptron and Hebbian 
learning algorithms for the case $a=2.0$. 
The AdaTron learning became the warst algorithm among the three.
}
\end{figure}

On the other hand, for the unlearnable case, 
the generalization error converges exponentially 
to $a$-dependent non-zero values 
both for perceptron 
and AdaTron learnings. 
Unfortunately, these residual 
errors are 
not necessarily the best possible 
value as seen in Fig. 2. 
From this figure, we see that for the 
unlearnable case the 
AdaTron learning is not superior to the perceptron learning, 
although the AdaTron learning is regarded as 
the most sophisticated learning 
algorithm for the learnable case \cite{Biehl}. 
In Fig. 6 we plot the generalization error of 
the perceptron, Hebbian and AdaTron learnings for the unlearnable case
($a=2.0$). 

For the Hebbian learning, 
the generalization error converges to $2H(a)$ for 
$a>a_{c1}=\sqrt{2{\log}2}$ and 
to $1-2H(a)$ for $a<a_{c1}$ as 
${\alpha}^{-1/2}$. 
For $a>a_{c1}$, this residual error $2H(a)$ 
corresponds to the optimal value. 
However, for $a<a_{c1}$, 
the generalization error 
of the Hebbian learning 
exceeds $0.5$ and, in addition, 
an over training is observed (Figs. 2, 3) 

This difficulty can be 
avoided partially by allowing 
the student to select 
suitable examples \cite{Kinzel}. 
If the student uses only examples which lie in 
the decision boundary, that is, if examples satisfy $u=0$,  
the generalization 
error converges to the optimal value as ${\alpha}^{-1/2}$ 
except only for $a_{c2}<a<a_{c1}$.
%%
%%
%%
%%%%%%%%%%%%%%%%%%%%%%%%%%%%%%%%%%%%%%%%%%%%%%%%%%%%%%%%%%%%%%%
%%%%%%%%%%%%%%%%%%%%%%%%%%%%%%%%%%%%%%%%%%%%%%%%%%%%%%%%%%%%%%%
\subsection{Optimization of learning rate}
%%%%%%%%%%%%%%%%%%%%%%%%%%%%%%%%%%%%%%%%%%%%%%%%%%%%%%%%%%%%%%%
%%%%%%%%%%%%%%%%%%%%%%%%%%%%%%%%%%%%%%%%%%%%%%%%%%%%%%%%%%%%%%%
%%
%%
We next regard the learning rate 
$g$ as a function of $\alpha$ and 
and construct an algorithm by optimizing $g$. 
In order to decide the optimal rate $g_{\rm opt}$ 
we maximize the right hand side of equation (\ref{dynamic}) with 
respect to $g$. 
This procedure is somewhat similar to 
the processes of determining 
the annealing schedule.
This optimization procedure is different from 
the method of Kinouchi and Caticha \cite{Kinouchi}. 

We apply this technique to the case of the perceptron, 
the Hebbian and the AdaTron learning algorithms. 
For the perceptron learning, this optimization procedure 
leads to the asymptotic form of generalization error 
as 
\begin{equation}
{\epsilon}_{g}=\frac{4}{\pi\alpha}
\end{equation} 
for the learnable case and to 
\begin{equation}
{\epsilon}_{g}=2H(a)+\frac{\sqrt{4{\pi}H(a)}}{{\pi}(1-2{\Delta})}
{\alpha}^{-1/2}
\end{equation} 
for the unlearnable case, where 
$2H(a)$ is the optimal value for $a>a_{c2}$. 
In the asymptotic region ${\alpha}{\rightarrow}\infty$, the 
learning rate $g_{\rm opt}$ behaves as 
$g_{\rm opt}\,{\sim}\,l/{\alpha}$. 
This learning strategy thus seems to work well 
for $a>a_{c2}$. 
However, at $a=a_{c1}$, 
this optimization procedure fails to 
reach the best possible value of 
the generalization error and the generalization 
ability deteriorates to $0.5$ (which is equal to 
the result by the random guess) \cite{Inoue1}. 
The reason is that for $a=a_{c1}$  
the optimal learning rate $g_{\rm opt}$ 
vanishes.

For the AdaTron learning, 
this type of 
optimization procedure 
gives the generalization ability as 
\begin{equation}
{\epsilon}_{g}=\frac{4}{3\alpha}
\end{equation}
for the learnable case and 
\begin{equation}
{\epsilon}_{g}=2H(a)+\frac{\sqrt{2}}{\pi}
\sqrt{\frac{2{\pi}H(a)+\sqrt{2\pi}a{\Delta}}
{4a^{2}{\Delta}}
}
\frac{1}{\sqrt{\alpha}}
\end{equation}
for the unlearnable rule.
Fortunately, for the AdaTron learning, 
the optimal learning late does not vanish 
even at $a=a_{c1}$, and 
therefore this optimization 
procedure works effectively for 
$a>a_{c2}$ \cite{Inoue2}. 

On the other hand, for the Hebbian learning,  
the above optimization procedure 
does not change the asymptotic form of 
the generalization error \cite{Inoue1}. 
Nevertheless, if we introduce the optimal 
learning rate $g_{\rm opt}$ into 
the Hebbian learning with queries, 
we get the very fast convergence of generalization error 
as 
\begin{equation}
{\epsilon}_{g}=2H(a)+\frac{\sqrt{c}}{\pi}{\exp}(-\frac{\alpha}{\pi}), 
\end{equation}
where $c$ is a positive constant. 

The present optimization procedure 
does not work effectively for 
$a<a_{c2}$ 
because the key point of this method consists 
in pushing the student toward the state $R=1$ and 
this state is not optimal 
for $a<a_{c2}$ (see Fig. 2).
%%
%%
%%
%%
%%
%%%%%%%%%%%%%%%%%%%%%%%%%%%%%%%%%%%%%%%%%%%%%%%%%%%%%%%%%%%%%%%%%%%%
%%%%%%%%%%%%%%%%%%%%%%%%%%%%%%%%%%%%%%%%%%%%%%%%%%%%%%%%%%%%%%%%%%%%
\section{Remarks}
%%%%%%%%%%%%%%%%%%%%%%%%%%%%%%%%%%%%%%%%%%%%%%%%%%%%%%%%%%%%%%%%%%%%
%%%%%%%%%%%%%%%%%%%%%%%%%%%%%%%%%%%%%%%%%%%%%%%%%%%%%%%%%%%%%%%%%%%%
%%
%%
In the present work, 
we have found that the  off-line learning obtain the best possible 
value of the generalization error for the whole range of $a$. 
On the other hand, the conventional on-line 
learning algorithm should be improved.
We could improve the conventional on-line learning strategies 
by introducing the time-dependent optimal learning rate, and 
queries. 
We could obtain the theoretical 
lower bound of the generalization error for the whole 
parameter range in the on-line mode. 
As our optimal learning rate contains 
the parameter $a$ unknown to the student, 
the result can be regarded only as a lower bound of 
the generalization error. 
However, if one uses the asymptotic form of $g_{\rm opt}$, 
the parameter independent learning algorithm 
can be formulated and the same generalization ability 
as the parameter dependent case 
can be obtained \cite{Inoue1,Inoue2}. 
%%
%%
%%

%%%%%%%%%%%%%%%%%%%%%%%%%%%%%%%%%%%%%%%%%%%%%%%%%%%%%%%%
We thank Professor Shun-ichi Amari 
for useful comments.  
J.I. is partially 
supported by the Junior Research Associate program of RIKEN. 
Y.K. is partially supported by a program ``Research 
for the Future (RFTF)'' of Japan Society for the 
Promotion of Science. 
And J.I. also acknowledges Professor C. Van den Broeck for 
stimulus discussions. 
 
%%%%%%%%%%%%%%%%%%%%%%%%%%%%%%%%%%%%%%%%%%%%%%%%%%%%%%%%%%
%%%%%%%%%%%%%%%%%%%%%%%%%%%%%%%%%%%%%%%%%%%%%%%%%%%%%%%%%%
%%
%%
%%%%%%%%%%%%%%%%%%%%%%%%%%%%%%%%%%%%%%%%%%%%%%%%%%%%%%%%%%%

%%%%%%%%%%%%%%%%%%%%%%%%%%%%%%%%%%%%%%%%%%%%%%%%%%%%%%%%%%%

\end{document}